\documentclass[conference]{IEEEtran}

\usepackage{tikz,psfrag,epsfig}
\input{tcilatex}

\begin{document}

\title{On the Network-Wide Gain of Memory-Assisted Source Coding}

\author{Mohsen Sardari, Ahmad Beirami, Faramarz Fekri\\
School of Electrical and
Computer Engineering, Georgia Institute of Technology, Atlanta, GA 30332\\
\texttt{Email:}\{mohsen.sardari, beirami, fekri\}@ece.gatech.edu

\thanks{This material is based upon work supported by the National Science Foundation under Grant No. CNS-1017234.}
}

\newcommand{\mc}{\mathcal}
\newcommand{\mb}{\mathbf}
\newcommand{\BH}{{\sf bit$\times$hop}}

\maketitle
\thispagestyle{empty}
\pagestyle{empty}
\begin{abstract}
Several studies have identified a significant amount of redundancy in the network traffic. For example, it is demonstrated that there is a great amount of redundancy within the content of a server over time. This redundancy can be leveraged to reduce the network flow by the deployment of memory units in the network. The question that arises is whether or not the deployment of memory can result in a fundamental improvement in the performance of the network.
In this paper, we answer this question affirmatively by first establishing the fundamental gains of memory-assisted source compression and then applying the technique to a network. Specifically, we investigate the gain of memory-assisted compression in random network graphs consisted of a single source and several randomly selected memory units. We find a threshold value for the number of memories deployed in a random graph and show that if the number of memories exceeds the threshold we observe network-wide reduction in the traffic.
\end{abstract}


\section{Introduction}
\label{sec:intro}
Several studies have demonstrated the existence of considerable amount of redundancy in the Internet data traffic, where a few major dimensions have been identified as the main sources of redundancy in the network traffic. 
For example, the contents of a web server contains more than $60\%$ redundant data on the average within a ten-day period~\cite{SIVA-TECH-REPORT}. Further, there are some popular files in each server that may be requested by several clients in the network. This redundancy in the data may be leveraged to reduce the communication within the network. 
The existing redundancy elimination techniques are mostly based on end-to-end caching mechanisms, where the redundant content is only cached in the server and the client for future reference~\cite{CLOUD-CONTROL}. However, these end-to-end approaches do not efficiently leverage the redundancy in the network because there is no memorization in today's Internet except end-to-end caching mechanisms~\cite{SIVA-TECH-REPORT}. 

Recently, a few studies considered the deployment of redundancy elimination techniques within the network~\cite{ anand_sigcomm_09,Siva_ACM}, where the intermediate nodes in the network have been assumed to be capable of caching of the previous communication and processing of the data. These works studied the network flow reduction via ad-hoc solutions such as deduplication of the repeated segments of the traffic without any connection to the information theory. The objective of this paper is to study this problem from an information theoretic point of view. We assume that some intermediate nodes (referred to as memory nodes) are capable of the memorization (to be defined later) of the previous communications which have passed through them.  We further assume that the memorized content will be used in a \emph{memory-assisted source coding} in the network, which we refer to as \emph{network flow compression with memory}.
However, several questions remain open regarding memory-assisted compression of the network flow. Does the deployment of memory in the network provide any fundamental benefit over end-to-end solutions? How much saving could be achieved using network compression? How can the savings be achieved?

This paper attempts to answer the above fundamental questions. To the best of our knowledge, this is the first work which addresses the memory-assisted redundancy elimination of the flow from the information theoretic point of view. 
The data that is transmitted inside the network have different spatial and temporal probability distributions. Thus, prior knowledge of the probability distributions underlying the contents may not be assumed. Hence, one important characteristic of any compression solution is that it must be \emph{universal} in the sense that it must be able to remove redundancy without knowing the statistics and nature of the data~\cite{LZ77, CTW95}. 

 In this paper, we focus our study on the fairly broad class of parametric information sources, which include the class of Markov sources of any finite order~\cite{IT11}. 
We formulate the problem as memory-assisted compression of the network flow, where the redundancy in the traffic data is to be removed.
In this context, the redundancy elimination could be viewed as universal source coding, where the goal is to represent the data with a minimum length codeword~\cite{IT11}. 
However, as we will discuss in Sec.~\ref{sec:problem_statement}, this problem is different from the distributed source coding problem because of memory units. We investigate the fundamental gain of the memory-assisted network flow compression over end-to-end universal compression techniques, where the content is compressed at the server and routed via routers without any memorization inside the network. 
Throughout this paper, we focus on the problem involving a \emph{single source} (content server) that is fixed in the network and extensions to multiple sources is left as future work.

In what follows, we first describe the memory-assisted source coding in Sec.~\ref{sec:problem_statement}. The memory deployment problem and the gain of memorization in networks is discussed in Sec.~\ref{sec:deployment}. In Sec.~\ref{sec:gain-on-randomgraph} we study memory-assisted network flow compression on Erd\H{o}s-R\'enyi (ER) random graphs and find a threshold value for the number of memory units. Finally, simulation results are provided in Sec.~\ref{sec:simulation}.

\section{Memory-Assisted Source Coding}
\label{sec:problem_statement}
\begin{figure}
\centering
\begin{tikzpicture}
\draw (0,0) node[circle, fill=blue!90, text=white](S){$S$};
\draw (1.5,0) node[circle, fill=blue!70, text=white](M){$\mu$};
\draw (3,.75) node[circle, fill=black!80, text=white](C1){\tiny{$C_1$}};
\draw (3,-.75) node[circle, fill=black!80, text=white](C2){\tiny{$C_2$}};

\draw [->, very thick] (S)	--	(M);
\draw [->, thick] (M)--	(C1);
\draw [->, thick] (M)--	(C2);	
\end{tikzpicture}
 \caption{The basic memory-assisted network flow compression scenario.} 
 \label{fig:single_hop}
\end{figure}
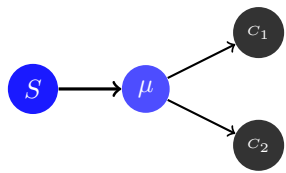

In what follows, we introduce the memory-assisted source coding via the sample network depicted in Fig~\ref{fig:single_hop}, which is consisted of a server $S$, a memory unit $\mu$ (which also acts as a router), and the clients $C_1$ and $C_2$. We assume that the communication is as follows. First, client $C_1$, who has not previously communicated with the server, acquires the sequence $f_1$ from the server through the intermediate node $\mu$. Next, client $C_2$, who also has not previously communicated with $S$, acquires the sequence $f_2$ from $S$ through $\mu$. 
In order to show the benefits of the deployment of memory in the router, we compare three schemes:
\begin{itemize}
\item {\em NcompNmem} ({No compression with no memory}), which does not apply any compression and does not utilize the memory unit.
\item {\em UcompNmem} ({Universal compression with no memory}), which only applies end-to-end universal source coding at the source without using the memory unit. 
\item {\em UcompWmem} (Universal compression with memory), which assumes that the router has memory and utilizes the memory unit when compressing the data at the source.
\end{itemize}
In all of the above scenarios, we assume that the client has no previous communication with the server since it is usually the case in networks. 
 On the other hand, the memory/router is capable of memorizing the communication of $f_1$ between $S$ and $C_1$ in order to better compress $f_2$ (on the link from $S$ to $\mu$) that is then being delivered to $C_2$.
We will later demonstrate that even if sequences $f_1$ and $f_2$ are {\em independent} given that the source model is {\em known}, the memorization of $f_1$ in $\mu$ can result in the reduction of the communication for the transfer of $f_2$ from $S$ to $C_2$. This seemingly counter intuitive phenomenon is due to the fact that the source model is {\em not} known a priori (at $\mu$). The underlying source coding must be universal, which imposes a compression overhead when the length of the sequence is finite (small)~\cite{IT11}. On the other hand, sequence $f_1$ does indeed contain some information about the unknown source parameters to the extent that an infinite length sequence $f_1$ can be used to identify all of the unknown parameters of the source.
This side information can be memorized at the memory unit $\mu$ and the source $S$ for the compression of $f_2$. Then the memory unit can decode $f_2$ using the side information and send $f_2$ to $C_2$. It is important to note that the saving of memory-assisted compression in terms of flow reduction is observed in the $S-\mu$ link. For example, if $f_1$ and $f_2$ are unit size and the memorization helps to compress $f_2$ by a factor of 2, the total flow is reduced from $1+1$, to $0.5+1$ \BH, where there is a gain $2$ in the link between $S$ and $\mu$.

While relevant, the memory-assisted source coding problem is different from those addressed by distributed source compression techniques (i.e., the Slepian Wolf problem) that target multiple correlated sources sending information to the same destination~\cite{slepian_wolf, pradhan:dcc99}. As described in the above example, the memory-assisted source coding gain is due to the fact that the source parameter is unknown. Therefore, when the length of the sequence $f_2$ increases, the memory-assisted source coding gain, with respect to UcomNmem, vanishes since the source parameter can be well estimated using the sequence, and hence, $f_2$ can be compressed to fundamental limit (i.e., entropy rate). On the other hand, in the Slepian-Wolf, the gains are achievable in the asymptotic regime. Further, the memorization of a sequence $f_1$ that is independent of the sequence $f_2$ can result in a gain in memory-assisted compression of $f_2$ whereas, in the Slepian-Wolf problem, the gain is due to the bit by bit correlation between the two sequences. 
The memory-assisted compression problem is also distinct from the lossless source coding techniques such as LZ and CTW that simply remove redundancy in a single piece of content without regard to any memorization at the intermediate node $\mu$~\cite{LZ77,CTW95}.

Next, we characterize the benefits of network memory in the context of a universal source coding problem for the class of smooth parametric sources~\cite{IT11,ISIT11}.
Let $l(C_n, x^n) = l_n(x^n)$  denote the length function that describes the codeword associated with the sequence $x^n$. 
 The expected codeword length $\mb{E} l_n(X^n)$ quantifies the compression performance of UcompNmem, where $\mathbf{E}[\cdot]$ denotes the expectation operator. For an asymptotically optimal code in the sense that it achieves the entropy rate, we have $\frac{1}{n} (\mb{E} l_n(X^n) - H_n(X^n))\to 0$ as $n \to \infty$, where $H_n(X^n)$ denotes the entropy of a sequence of length $n$.
In the case that the router is capable of memorization (UcompWmem), assume that $y^m$ is another sequence of length $m$ from the same source that generates $x^n$.
By context memorization, we mean that both the source $S$ and the memory unit $\mu$ have already visited the sequence $y^m$. Let $l_{n|m}$ be a regular length function where $S$ and $\mu$ have access to a context memory of length $m$ from the previous communications. Then, the expected codeword length with memory $\mb{E}l_{n|m}(X^n)$ characterizes the compression performance of UcompWmem for $x^n$ in the link from $S$ to $\mu$, in Fig~\ref{fig:single_hop}.

Let $Q(l_n, l_{n|m})$ be defined as the ratio of the expected codeword length of UcompNmem to that of UcompWmem as
\begin{equation}
Q(l_n, l_{n|m}) \triangleq \frac{\mb{E} l_n(X^n)}{\mb{E} l_{n|m}(X^n)}.
\end{equation}
Further, let $\epsilon$ be a real number such that $0<\epsilon<1$. We denote $g(n,m, \epsilon)$ as  the fundamental gain of the context memorization on the family of parametric sources on a sequence of length $n$ using a context sequence of length $m$ for a fraction $1-\epsilon$ of the sources, which is defined as follows:
\begin{equation}
g(n,m,\epsilon) = \sup_{z \in \mathbb{R}} \left\{z : ~\mb{P}\left[Q(l_n, l_{n|m})\geq z\right] \geq 1-\epsilon\right\}.
\label{eq:g_definition}
\end{equation}
In other words, the fundamental gain of memorization is at least $g(n,m,\epsilon)$ for a fraction $1-\epsilon$ of the sources in the family.

In~\cite{ISIT11}, Beirami and Fekri studied the fundamental limits of compression without memory. In~\cite{IT11}, they extended their study to the
behavior of the memorization gain $g(n,m,\epsilon)$ with respect to the sequence length and the memory size for different source models. In the rest of this paper, we investigate the effect of the context memorization gain on a network given that the memory-assisted compression gain $g(n,m,\epsilon)$ (hereafter, referred as $g$) is known. 

\section{Memory Deployment Problem in Networks}
\label{sec:deployment}
In this section, we will investigate the memorization gain on a network. A network is represented by an undirected graph $G(V,E)$ where $V$ is the set of $N$ nodes (vertices) and $E=\{uv:u,v \in V\}$ is the set of edges connecting nodes $u$ and $v$. We consider a single source $S$ which is the content server, and a set of memories $\mathbf{\mu} = \{ \mu_i\}_{i=1}^M$ chosen out of $N$ nodes. The content server is assumed to be a parametric information source~\cite{IT11}. We assume that each client requests a small to moderate length size sequence from the content server. As discussed in Sec.~\ref{sec:problem_statement}, there is a fundamental limit beyond the entropy on the universal compression of a small to moderate length sequence. Therefore, UcompNmem will only be able to compress the content to a value which may be significantly larger than the entropy of the sequence. On the other hand, a memory nodes $\mu_i$ is capable of memorizing the communication between the server and some client node $C_i$.

To investigate the gain of memorization in the compression of the network flow, we must consider two phases. The first is the memorization phase in which we may assume all memory units have visited some sufficiently long sequence (or equivalently, a sufficient collection of small to moderate length sequences) from  the source. This phase is realized in actual communication networks by observing the fact that a sufficient number of clients may have previously retrieved small to moderate length sequences from the server such that, via their routing, each of the memory units has been able to memorize the source. In the second phase, which is the subject of this section, we assume each node in the entire network may request (a small to moderate length) content from the server, uniformly. Our goal is to characterize the memorization gain in the compression of the sequences that are retrieved in the second phase.
The above view simplifies our study as we are not concerned with the transition phase during which the memorization is taking place in the memory units. Hence, we can assume each memory unit will provide the same memory-assisted compression gain of $g$ of the link from source to itself that depends on the length of the sequence that is being transmitted as well as the length of the sequence that is memorized in the memory unit, as described in Sec.~\ref{sec:problem_statement}.

The goal of the memory deployment is to minimize the total cost of communication between the source and destinations in the network, measured by \BH, by deploying a set of memories $\mathbf{\mu}$. We wish to study the behavior of the total savings in terms of \BH, as a function of the number of the memories, place of memories, size of memories, length of the sequences, and the information source model.

In a network with source $S$ and a set of destinations $\mathbf{D}= \{D_i\}_{i=1}^N$, let $f_D$ be the flow destined to $D \in \mathbf{D}$. The distance between any two nodes $u$ and $v$ is shown by $d(u,v)$. The distance is measured as the number of hops in the shortest path between two nodes. As we will see later, introducing memories to the network will change the lowest cost paths from the source to destinations, as there is a gain associated with the $S-\mu$ portion of the path. Therefore, we have to modify paths accounting for the gain of memories. Accordingly, for each destination $D$, we define \emph{effective walk}, denoted by $W_D = \{S,u_1,\ldots,D\}$, which is the ordered set of nodes in the modified (lowest cost) walk between the source and $D$. Then, we partition the set of destinations as $\mathbf{D} = \mathbf{D}_1 \cup \mathbf{D}_2$, where $\mathbf{D}_1=\{D_i:\exists {\mu}_{\scriptscriptstyle D_i} \in W_{D_i}\}$ is the set of destinations observing a memory in their effective walk, and $\mu_{\scriptscriptstyle D_i} = \arg \min_{\mu \in \mathbf{\mu}} \{\frac{d(S,{\mu})}{g} + d({\mu}, {\scriptstyle D_i})\}$. The total flow $\mathcal{F}$ is then defined as
\begin{equation}
\label{eq:totalflow}
\begin{array}{lcl}
\mathcal{F} &=&
\sum_{D_i \in \mathbf{D}_1}\left(\frac{f_{D_i}}{g}d(S,{\mu}_{\scriptscriptstyle {D_i}}) + f_{\scriptscriptstyle D_i}d({\mu}_{\scriptscriptstyle D_i},{\scriptstyle D_i} ) \right) +\\
&&\sum_{D_j \in \mathbf{D}_2} f_{D_j} d(S,D_j)
\end{array}						
.\end{equation}
Using~\eqref{eq:totalflow}, we define $\hat{d}_D$, called the \emph{effective distance} from source to $D$, as
\begin{equation}
\label{eq:effective_dist}
\hat{d}_D = \left\{
\begin{array}{ll}
\frac{d(S,{\mu}_{\scriptscriptstyle D})}{g} + d({\mu}_D, {\scriptstyle D})	& D\in  \mathbf{D}_1\\
d(S,D) 							& D\in  \mathbf{D}_2
\end{array}
\right.
.\end{equation}
In short, the effective distance is the distance when memory-assisted source compression is performed and hence the gain $g$ applies. By definition, $\hat{d}_D \leq d(S,D)~ \forall D$.
For simplicity, we assume $f_D=1$ for all destinations.
Hence,
$
\mathcal{F} = \sum_{D \in \mathbf{D}}\hat{d}_D 
.$

In a general network where every node can be a client, we define a generalized network-level gain of memory deployment as a function of memorization gain $g$, as follows:
\begin{equation}
\mathcal{G}(g)=\frac{\mathcal{F}_0}{\mathcal{F}}
,\end{equation}
where $\mathcal{F}_0$ is the total flow in the network under UcompNmem, i.e., $\mathcal{F}_0 = \sum_{D\in \mathbf{D}} d(S,D)$.

In order to show the challenges of the memory deployment problem, we show as to how a single memory changes the effective paths in a network with a single source.
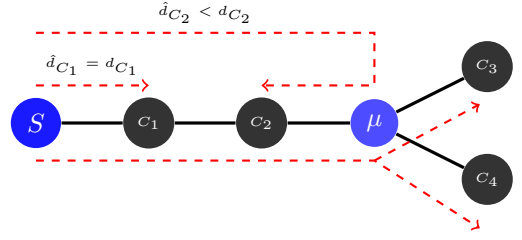
\begin{figure}
\centering
\begin{tikzpicture}
\draw (0,0) 	node[circle, fill=blue!90, text=white]	(S){$S$};
\draw (1.5,0) 	node[circle, fill=black!80, text=white]	(C1){\tiny{$C_1$}};
\draw (3,0) 	node[circle, fill=black!80, text=white]	(C2){\tiny{$C_2$}};
\draw (4.5,0) 	node[circle, fill=blue!70, text=white]	(m){{$\mu$}};
\draw (6,.75) 	node[circle, fill=black!80, text=white]	(C3){\tiny{$C_3$}};
\draw (6,-.75) 	node[circle, fill=black!80, text=white]	(C4){\tiny{$C_4$}};

\draw [very thick] (S)	--	(C1);
\draw [very thick] (C1)	--	(C2);
\draw [very thick] (C2)	--	(m);
\draw [very thick] (m)	--	(C3);
\draw [very thick] (m)	--	(C4);

\draw [->, red, thick, dashed] (0,.5)	--	(1.5,.5) node[midway, above, text=black] {\tiny{$\hat{d}_{C_1}={d}_{C_1}$}};
\draw [->, red, thick, dashed] (0,-.5)	-- (4.5, -.5)  --	(5.9,-1.4); 
\draw [->, red, thick, dashed] (4.5, -.5) to (5.9,0.25);

\draw [->,red,thick, dashed] (0,1.2) --  (4.5,1.2)  node[midway, above, text=black] {\tiny$\hat{d}_{C_2}<{d}_{C_2}$} --  (4.5,0.5) to (3,.5) ;
\end{tikzpicture}
\caption{The shortest walk between the source and destination is not necessarily a path when we have memory in network.}
\label{fig:line}
\end{figure}
Consider the network with the source node $S$ placed as shown in Fig.~\ref{fig:line}. The destinations are nodes $C_1,\ldots,C_4$, and $g=4$. The effective walks from the source to destinations are obviously the shortest paths when there is no memorization (UcompNmem). As shown in the figure, the placement of memory changes the effective path to $C_2$ while the shortest paths from the source to $C_1$, $C_3$, and $C_4$ are the same as the effective paths.
Without memory, the shortest path to $C_2$ is two hops long ($S\rightarrow C_1 \rightarrow C_2$), while the memory totally changes the effective walk distance to $C_2$ to $\hat{d}_{C_2} = \frac{3}{4}+1$ as depicted in the figure.

In order to extend our study to analyze the achievable gains in general networks and study the behavior of $\mathcal{G}(g)$, in the next section, we consider the memory deployment gain in the network graphs that resemble the ER random graph~\cite{erdos}. The ER random graph is the building block of the recent models for complex graphs and hence the results would be useful in much broader contexts. We specifically direct our attention to connected random graphs since they better describe real networks.

\section{Gain of Memory Deployment on the Network Flow Compression}
\label{sec:gain-on-randomgraph}

\begin{definition}
An ER random graph $G(N,p)$ is an undirected, unweighted graph on $N$ vertices where every two vertices are connected with probability $p$.
\end{definition}
\begin{definition}
Let $u,v \in G$ be any two vertices. The diameter of a connected graph is defined as $\max_{u,v} d(u,v)$. Similarly, the average distance of a connected graph is defined as $\mathbf{E}[d(u,v)]$.
\end{definition}

The following properties hold for ER random graphs:
\begin{enumerate}
  \item  If $p<\frac{(1-\epsilon)\log N}{N}$, then $G(N,p)$ almost surely (a.s.) has isolated vertices and thus disconnected.
  \item  If $p=\frac{c\log N}{N}$ for some constant $c>1$, then $G\left(N,p\right)$ is a.s. connected and every vertex asymptotically has degree $c\log N$~\cite{alon2008}.
  \item The diameter of $G(N,p)$ is almost surely $\frac{\log N}{\log Np}$.
  \item The average distance in $G(N,p)$, denoted by $\bar{d}$, is
  \begin{equation}
\label{eq:avg_dist}
\bar{d} = (1+o(1))\frac{\log N}{\log Np},
\end{equation}
provided that $\frac{\log N}{\log Np}$ goes to infinity as $N \rightarrow \infty$ (this condition is satisfied in the connected regime)~\cite{chung-book}.%
\end{enumerate}

Again, the main question is how $\mathcal{G}(g)$ scales with $M$. In order to characterize the gain of memory placement, we consider connected $G(N,p), p=\frac{c\log N}{N}$, with a single source node $S$ and all other nodes as destinations. Since the expected degree of all nodes in ER graph is the same and every vertex is a destination with equal probability, we select memories $\{\mu_i\}_{i=1}^M$ uniformly and random. Theorem~\ref{thm:scaling}, below, provides the scaling of $\mathcal{G}(g)$ with respect to $M$:

\begin{theorem}
\label{thm:scaling}
Suppose $M$ is the number of deployed memories in an ER random graph. Let $\epsilon$ be a positive real number.
\begin{itemize}
  \item  [(a)] If $M = O\left(N^{\frac{1}{g}-\epsilon}\right)$, then $\mathcal{G}(g)\sim 1$. \footnote{In this paper, we have used the following asymptotic notation: $f(x) \sim g(x)$ iff $f(x)/g(x) \rightarrow 1$.}
  \item  [(b)] If $M = \Omega\left(N^{\frac{1}{g}+\epsilon}\right)$, then all the destinations benefit from memory and $\mathcal{G}(g) \stackrel{a.s.}{=}  \frac{g}{1-g\log_N(\frac{M}{N})}$.
\end{itemize}
\end{theorem}
\begin{IEEEproof}[Sketch of the proof]
We first find an upper bound on the number of destinations benefit form each memory. This upper bound is sufficient to derive part (a) of the theorem. For the second part, we find a lower bound on the number of benefiting destinations. 
\end{IEEEproof}

To characterize $\mathcal{G}(g)$, we first need to find $\mathcal{F}_0$. The average distance from the source to a node is $\bar{d}$. Thus, $\mathcal{F}_0 = N\bar{d}$. For large $N$, \eqref{eq:avg_dist} results in
$
\mathcal{F}_0 \sim  \frac{N\log N}{\log \log N}
.$

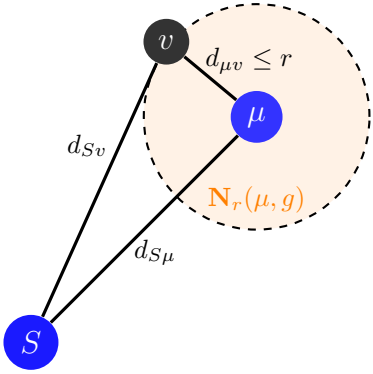
\begin{figure}
\centering
\begin{tikzpicture}
\draw (0,0) 	node[circle, fill=blue!90, text=white]	(S){\large{$S$}};
\draw [thick, dashed, fill=orange!10, text=orange] (3,3) circle (1.5) (3,1.9) node {$\mathbf{N}_r(\mu,g)$};
\draw (3,3) 	node[circle, fill=blue!80, text=white]	(m){\large{$\mu$}};

\draw (1.8,4) 	node[circle, fill=black!80, text=white]	(v){\large{$v$}};

\draw [very thick] (S)	--	(m) (1.65,1.2) node {$d_{S\mu}$};
\draw [very thick] (m)	--	(v) (2.9,3.75) node {$d_{\mu v}\leq r$} ;
\draw [very thick] (S)	--	(v) (.75,2.62) node {$d_{Sv}$} ;

\end{tikzpicture}
\caption{Memory Neighborhood}
\label{fig:neighbor}
\end{figure}

Next, we need to find $\mathcal{F}$. For every memory $\mu$ we consider a neighbourhood $\mathbf{N}_r(\mu)$ as shown in Fig.~\ref{fig:neighbor}. This neighborhood consist of all vertices $v$ within distance $r$ from $\mu$. We choose $r$ such that, almost surely, all nodes in $\mathbf{N}_r(\mu)$ would benefit from the memory $\mu$. Clearly, if $\frac{d(S,\mu)}{g} + r = d(S,v)$, the benefit of the memory for node $v$ vanishes and only nodes at distances less than $r$ benefit from the memory $\mu$. Given $g$, we denote this set of nodes benefiting from $\mu$ by $\mathbf{N}_r(\mu,g)$.
\begin{equation}
\label{eq:neibor}
\mathbf{N}_r(\mu,g) = \left\{v:\frac{d(S,\mu)}{g} + d(\mu, v) \leq d(S,v)\right\}
.\end{equation}
Since memories are uniformly placed, the average value of $d(S,\mu)$ in $\hat{d}_v$ is equal to $\bar{d}$. Similarly, the average of $d(S,v)$ is also $\bar{d}$. Hence, solving for $r$ in~\eqref{eq:neibor} and then using the result on the average distance in~\eqref{eq:avg_dist}, we conclude
\begin{equation}
\label{eq:neigh-radius}
r \stackrel{a.s.}{=} (1-{1}/{g})\left(\frac{\log N}{\log \log N}\right) .
\end{equation}
The following lemma, by Chung and Lu~\cite{chung-book}, gives an upper bound on the total number of vertices in the neighborhood $\left |\mathbf{N}_r(\mu_i,g)\right |$, where $|\cdot|$ is the set size operator.
\begin{lemma}[\cite{chung-book}]
\label{lem:neighbor}
Assume a connected random graph. Then, for any $\epsilon>0$, with probability at least $1-\frac{1}{(\log N)^2}$, we have
$ \left |\mathbf{N}_r(\mu_i,g)\right | \leq (1+2\epsilon)(Np)^r$, for $1\leq r \leq \log N$.
\end{lemma}

Using Lemma~\ref{lem:neighbor} and~\eqref{eq:neigh-radius}, we deduce that
\begin{eqnarray}
\label{eq:gain-vertex}
|\mathbf{N}_r(\mu_i,g)| &\stackrel{a.s.}{\leq}&   (1+2\epsilon)(\log N)^{(1-\frac{1}{g})\left(\frac{\log N}{\log \log N}\right)} \nonumber\\
&=& (1+2\epsilon)N^{1-1/g}
.\end{eqnarray}
Therefore, the total number of nodes gaining from the memories is upper-bounded by
$
\sum_{i=1}^M |\mathbf{N}_r(\mu_i,g)| \leq M(1+2\epsilon)N^{1-1/g}
.$
As we will see, from~\eqref{eq:gain-vertex} it is clear that the gain of memory vanishes if $M$ is chosen small. The value $N^{1/g}$ is the threshold value for the network-wide gain. More accurately, if $M = O\left(N^{\frac{1}{g}-\epsilon}\right)$, there is no gain from memories.

\begin{IEEEproof}[Proof of Theorem~\ref{thm:scaling}(a)]
For all the nodes in $\mathbf{N}_r(\mu_i,g)$, we have a flow gain of $g$. Let $M = N^{\frac{1}{g}-\epsilon}$, then we have
\begin{eqnarray}
	\label{eq:double_counting}
\mathcal{G}(g)
&\leq& 		\frac{N\bar{d}}{\frac{\bar{d}}{g}M|\mathbf{N}_r(\mu,g)|+\bar{d}(N-M|\mathbf{N}_r(\mu,g)|)}\\
	\label{eq:insert-neighbor-size}
&\stackrel{a.s.}{\leq}& 		\frac{N}{N-(1-1/g)M  N^{(1-\frac{1}{g})}} \\
&=&			\frac{N}{N-(1-1/g)N^{1-\epsilon}} \sim		1 \nonumber
,\end{eqnarray}
where inequality in~\eqref{eq:double_counting} follows from the double counting  of the destination nodes that may reside in more than one neighborhood. Also,~\eqref{eq:insert-neighbor-size} follows from replacing~\eqref{eq:gain-vertex} in~\eqref{eq:double_counting}.
\end{IEEEproof}

Since we need more than $ n^{\frac{1}{g}}$ memory units to have a network-wide gain, the next question is as to how $\mathcal{G}(g)$ scales when the number of memory units exceeds $ n^{\frac{1}{g}}$. To answer this question, we need to establish a lower-bound on the neighborhood size and the number of nodes benefiting from memory. Further, we have to account for the possible double counting of the intersection between the memory neighborhoods. We use the following concentration inequality from~\cite{chung-book} to establish the desired bound.
\begin{proposition}[\cite{chung-book}]
\label{thm:concetration}
If $X_1,X_2,\ldots,X_n$ are non-negative independent random variables, then the sum $X=\sum_{i=1}^n  X_i$ holds the bound
\[
\mathbf{P}[X\leq \mathbf{E}[X]-\lambda]\leq \exp\left({-\frac{\lambda^2}{2\sum\mathbf{E}[X_i^2]}}\right)
.\]
\end{proposition}
This inequality will be helpful to show that the quantities of interest concentrate around their expected values.

The following lemma provides a lower-bound on the neighborhood size $\left |\mathbf{N}_r(\mu,g)\right |$ and the lower-bound on $\mathcal{G}(g)$, as we show, is immediate.
\begin{lemma}
\label{lem:lowerbound-neighbor}
Consider a set of vertices $V$ of $G(N,p)$ such that $\frac{|V|}{N}=o(1)$. For $0<\epsilon<1$, with probability at least $1-e^{-{Np|V|}{\epsilon^2}/2}$, we have
\begin{equation}
\label{eq:neighbor-lowerbound}
\left |\mathbf{N}_r(\mu,g)\right | \geq (1-\epsilon)(Np)^r.
\end{equation}
\end{lemma}
\begin{IEEEproof}
The vertex boundary of $V$, denoted by $\Gamma(V)$, consists of all vertices in $G$ adjacent to some vertex in $V$.
\[
\Gamma(V)=\left \{u:u\not \in V, \text{ and }u \text{ is adjacent to }v\in V\right\}
.\]
Let $X_u$ be the indicator random variable that a vertex $u$ is in $\Gamma(V)$, i.e., $\mathbf{P}[X_u=1] = \mathbf{P}[u\in\Gamma(V)]$. Then,
\begin{eqnarray}
\mathbf{E}\left[|\Gamma(V)|\right] &=& \sum_{u\not\in V} \mathbf{E}[X_u] = \sum_{u\not\in V}\mathbf{P}[u\in\Gamma(V)]\nonumber\\
&=&	\sum_{u\not\in V}\left(1-(1-p)^{|V|}\right) \nonumber \\
\label{eq:expectedboundary-1}
&\geq& p|V|(N-|V|) = (1-o(1))Np|V|
\end{eqnarray}
where the inequality in~\eqref{eq:expectedboundary-1} follows from
\[
\mathbf{P}[u\in\Gamma(V)]	=	1-(1-p)^{|V|} \geq 1-e^{-p|V|} \approx p|V|
,\]
and the second part holds because $\frac{|V|}{N}=o(1)$. Since, $X_u$'s are non-negative independent random variables, by applying Proposition~\ref{thm:concetration} with $\lambda = \sqrt{\alpha\mathbf{E}[|\Gamma(V)|]}$, with probability at least $1-e^{-\alpha/2}$ we have
\begin{eqnarray}
\label{eq:gamma-size}
|\Gamma(V)| &\geq& \mathbf{E}\left[|\Gamma(V)|\right] - \sqrt{\alpha\mathbf{E}[|\Gamma(V)|]}\nonumber\\
&\geq& (1-\epsilon)Np|V|
.\end{eqnarray}
By picking a single vertex and applying~\eqref{eq:gamma-size} inductively $r$ times, and then adding up the number of adjacent nodes, we obtain~\eqref{eq:neighbor-lowerbound}.
\end{IEEEproof}

Now that we have a lower-bound on the number of nodes benefiting from each memory, we show that by increasing the number of memories beyond $M=N^{\frac{1}{g}}$, memories cover all the nodes in the graph effectively and hence all the nodes would gain from the memory placement.

In order to limit the intersection between the neighborhoods, we reduce $r$ to $r_\delta$ as below:
\begin{equation}
\label{eq:r-delta}
r_\delta = (1-{1}/{g}-\delta)\left(\frac{\log N}{\log \log N}\right) .
\end{equation}
With this choice of $r_\delta$, by lemmas~\ref{lem:neighbor} and~\ref{lem:lowerbound-neighbor}, we deduce that the probability that a random node $u\in G$ belongs to the neighborhood $\mathbf{N}_{r_\delta}(\mu_i,g)$ of the memory $\mu_i$ is $N^{-1/g-\delta}$. Hence, the expected number of the covered nodes is
\begin{eqnarray}
\label{eq:neighbor-delta}
\mathbf{E}\left[\left|\bigcup_{i=1}^M \mathbf{N}_{r_\delta}(\mu_i,g)\right|\right]	&=& \sum_{u\in G}\mathbf{P}\left[u \in \cup_{i=1}^M \mathbf{N}_{r_\delta}(\mu_i,g)\right] \nonumber\\
&=&	\sum_{u\in G}\left(1-(1-N^{-1/g-\delta})^M\right) \nonumber\\
&\approx& N\left(MN^{-1/g-\delta}\right)= N
,\end{eqnarray}

where~\eqref{eq:neighbor-delta} holds by choosing $M=N^{1/g+\delta}$.

To show that the number of covered nodes is concentrated around its mean, we use Prop.~\ref{thm:concetration} again with $\lambda = \sqrt{\alpha\mathbf{E}\left[\left|\cup \mathbf{N}_{r_\delta}(\mu_i,g)\right|\right]}$. Then, with probability at least $1-e^{-\alpha/2}$ we have
\begin{eqnarray*}
\left|\bigcup_{i=1}^M \mathbf{N}_{r_\delta}(\mu_i,g)\right|	&\geq& \mathbf{E}\left[\left|\cup \mathbf{N}_{r_\delta}(\mu_i,g)\right|\right]- \lambda\\
&\geq&	(1-o(1))N
.\end{eqnarray*}
Hence, the memories cover, almost surely, all of the nodes.


\begin{figure}
\centering
{
\includegraphics[width = .9\columnwidth]{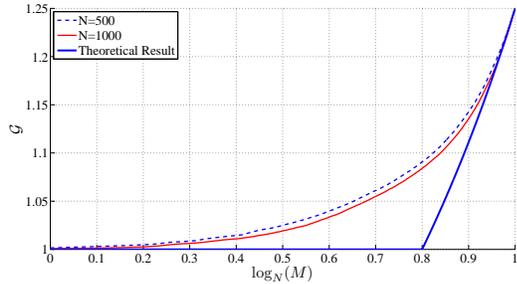}
\caption{Network-level gain $\mathcal{G}$ versus $\log_N(M)$ for different network sizes $N$ and $g=1.25$.}
\label{fig:simulation}
}
\end{figure}

Since all nodes are covered with high probability, we can associate each node with a neighborhood $|\mathbf{N}_{r_\delta}(\mu_i,g)|$, for which nodes' distances in the neighborhood from memory are $(1-o(1))r_\delta$.
\begin{IEEEproof}[Proof of Theorem~\ref{thm:scaling}(b)]
 By~\eqref{eq:neighbor-delta}, we can bound the network-wide gain of the memory from below. We have
\begin{eqnarray}
\label{eq:gain-lowerbound-as}
\mathcal{G}(g) &\stackrel{a.s.}{=} & \frac{N\bar{d}}{({\bar{d}}/{g}+r_\delta)N} \\
\label{eq:gain-lowerbound}
&=& 	\frac{1}{{1}/{g}+(1-{1}/{g}-\delta)} =	\frac{1}{1-\delta}
,\end{eqnarray}
where~\eqref{eq:gain-lowerbound-as} holds because the distance of the nodes from memory is $r_\delta$, asymptotically almost surely.
\end{IEEEproof}
As the number of memories becomes close to $N$, i.e., $\delta \rightarrow (1-\frac{1}{g})$, the gain $\mathcal{G} \rightarrow g$, as expected. In the next section, we verify our result in memory-assisted source coding and the network-wide gain of memory via numerical simulations.

\section{Simulation Results}
\label{sec:simulation}
In this section, we first demonstrate our theoretical results through an example. We consider the source to be a first-order Markov source with alphabet size equal to $256$. In~\cite{IT11}, Beirami and Fekri derived a lower bound on the memorization gain as a function of the sequence length and the memory size.
For example, they showed that $g(512$kB$,8$MB$,0.05)$ is about $1.25$, i.e., with a memory of $8$MB, a gain of $1.25$ is obtained on the memory-assisted compression of $512$kB long sequences~\cite{IT11}.
Fig.~\ref{fig:simulation} presents the simulation results for network-wide gain for different network sizes versus $\log_N(M)$ when $g=1.25$. The rightmost solid curve is our theoretical result in~\eqref{eq:gain-lowerbound-as}. For small values of $M$, the network-wide gain would be 1 for $N\rightarrow\infty$, while for large $M$, $\mathcal{G}$ tends to $g$. Also, as $N$ increases, simulation results approach the theoretical limit for both small and large values of $M$.


\bibliographystyle{IEEEtran}
\bibliography{net-comp}
\end{document}